\documentclass[letterpaper,english,prl,twocolumn]{revtex4}
\usepackage[T1]{fontenc}
\usepackage[latin9]{inputenc}
\usepackage{verbatim}
\usepackage{amsmath}
\usepackage{amssymb}
\usepackage{graphicx}
\usepackage{dsfont}
\usepackage{float}
\usepackage{cancel}
\usepackage{scrextend}
\usepackage{soul}
\usepackage{overpic}
\usepackage{wasysym}
\usepackage{braket}

\makeatletter

\pdfpageheight\paperheight
\pdfpagewidth\paperwidth

\usepackage{color}
\usepackage[usenames,dvipsnames,svgnames,table]{xcolor}
\usepackage{comment}
\usepackage{ulem}

\definecolor{violet}{rgb}{0.58, 0.0, 0.83}

\newcommand{\gae}{\lower 2pt \hbox{$\,
\buildrel{\scriptstyle >}\over {\scriptstyle \sim}\,$}}
\newcommand{\lae}{\lower 2pt \hbox{$\,
\buildrel{\scriptstyle <}\over {\scriptstyle \sim}\,$}}

\usepackage{graphicx}
\usepackage{color}

\begin{document}

\title{Inverted many-body mobility edge in a central
qudit problem}
\author{Saeed Rahmanian Koshkaki, Michael H. Kolodrubetz}
\affiliation{Department of Physics, University of Texas at Dallas, Richardson, TX, USA}

\begin{abstract}
    Many interesting experimental systems, such as cavity QED or central spin models, involve global coupling to a single harmonic mode. Out-of-equilibrium, it remains unclear under what conditions localized phases survive such global coupling. We study energy-dependent localization in the disordered Ising model with transverse and longitudinal fields coupled globally to a $d$-level system (qudit). Strikingly, we discover an inverted mobility edge, where high energy states are localized while low energy states are delocalized. Our results are supported by shift-and-invert eigenstate targeting and Krylov time evolution up to $L=13$ and $18$ respectively. We argue for a critical energy of the localization phase transition which scales as $E_c \propto L^{1/2}$, consistent with finite size numerics. We also show evidence for a reentrant MBL phase at even lower energies despite the presence of strong effects of the central mode in this regime. Similar results should occur in the central spin-$S$ problem at large $S$ and in certain models of cavity QED.
\end{abstract}

\maketitle

\begin{figure}[b]
\centering
\includegraphics[width=0.94\linewidth]{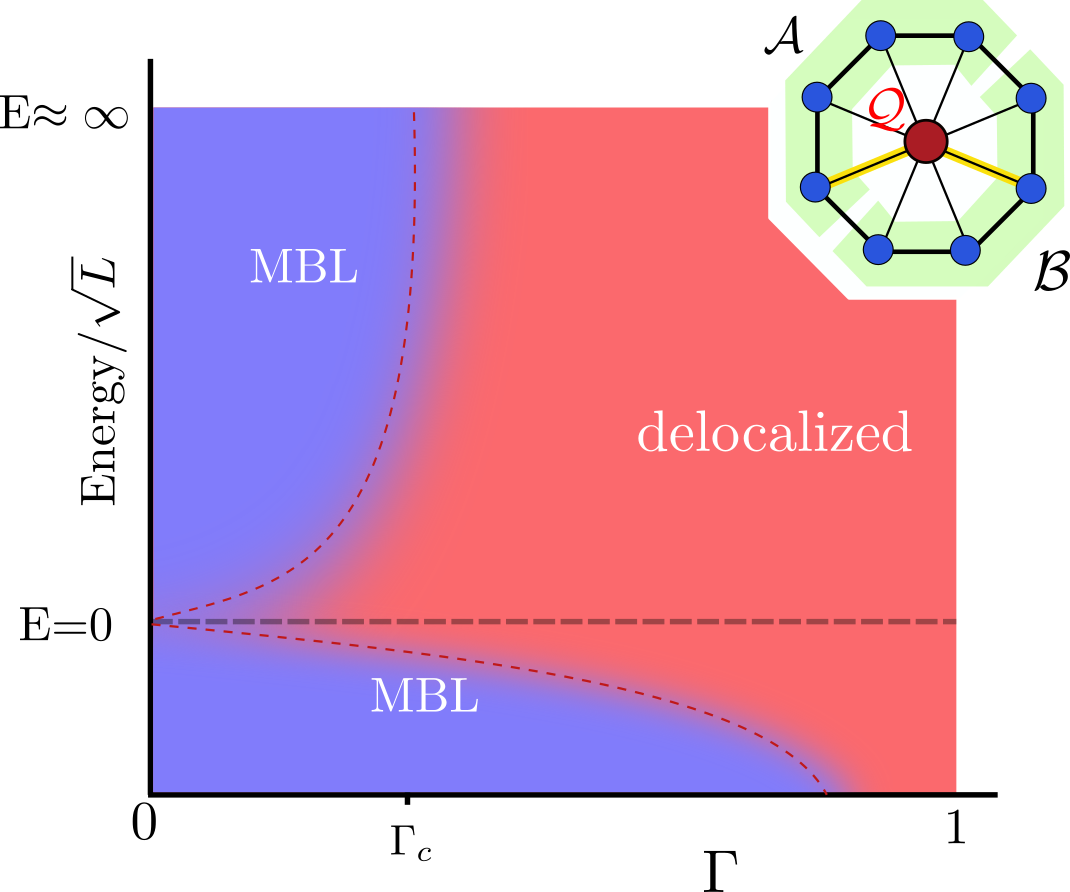}
\caption{Proposed phase diagram of inverted mobility edge for Ising model in the presence of  global qudit or spin-$S$ mode. For $\Gamma < \Gamma_c \approx 0.33$, we predict a delocalized to MBL phase transition as the energy is increased -- an inverted mobility edge.}
\label{fig:PhaseDia}
\end{figure}

Improvements in quantum control have brought non-equilibrium quantum systems to the forefront of condensed matter and AMO physics. Novel phases of matter are possible out of equilibrium, most of which require many-body localization \cite{WoottonMBLorder,HuseLocalizedOrder}. Many-body localization (MBL) results when sufficiently strong disorder prevents ergodicity in interacting systems, and is the only known generic route to avoid thermal equilibrium in isolated quantum systems \cite{NandkishoreMBLreview, Abanincolloquia}. Most numerical and analytical claims of MBL rest upon the assumption of low-dimensional, locally interacting Hamiltonians, and sufficiently long-range non-confining interactions are generally believed to destroy MBL \cite{NandkishoreMBLlongrange, YaoMBLLongrange, PonteCentralSpin,Hetterich2018}.

It was therefore surprising when we recently found that MBL can survive coupling to a global degree of freedom \cite{NathanMikeCentral}. Global coupling to a photon is a common occurrence in many-body cavity QED, where the cavity mode is primarily used to create all-to-all interactions between the atoms. The key result of \cite{NathanMikeCentral} was that the strength of this interaction is controlled by photon number in the cavity, $N$. If one takes the number of atoms $L$ to infinity while keeping the ratio $N/\sqrt{L}$ fixed, all-to-all interactions remain sufficiently weak to allow an MBL phase.

This opens the interesting possibility that, as the photon number -- or equivalently the energy -- is lowered, all-to-all interactions will reemerge and thermalize the system. This implies localization at high energies and thermalization at low energies, leading to an inversion of the conventional many-body mobility edge. In this paper, we will confirm that hypothesis using numerical and analytical tools, further uncovering a reentrant MBL phase at even lower energies. While similar phenomena occur in cavity QED, we argue that they are more favorable in non-bosonic models such as the central spin-$S$ and central $d$-level system (qudit).

\emph{Model} -- 
We start from the same Hamiltonian as \cite{NathanMikeCentral}, which was motivated by a standard model of spin-1/2 particles undergoing Floquet many-body localization \cite{ZhangVedikaHuse}. In the Floquet extended zone picture, the time-periodic drive is treated quantum mechanically by mapping it to a harmonic mode. This is represented geometrically in the inset to Figure \ref{fig:PhaseDia}. The spins form a locally coupled chain with periodic boundary conditions. These spins all couple globally to a single degree of freedom, such as a cavity photon or central spin-$S$. The goal of this work will be to study the low energy limit, where quantization of the central degree of freedom becomes important.

Specifically, our Hamiltonian can be written
\begin{equation} \label{Hez}
H = \frac{H_+}{2} + \frac{H_-}{4} (\hat{a}+\hat{a}^{\dagger}) + \hat{n} \Omega,
\end{equation}
where $H_\pm = H_z \pm H_x$, 
\begin{eqnarray}
   H_x & = & \sum_{i=1}^{L} g \Gamma \sigma^x_i~,\nonumber \\
   H_z & = & \sum_{i=1}^{L}\sigma_i^z\sigma_{i+1}^z + \sum_{i=1}^{L}(h+g\sqrt{1-\Gamma^2}G_i)\sigma_i^z~, \nonumber \\
   \hat{a}^{\dagger} & = &  \sum_{n=1}^{d-1} |n\rangle \langle n-1|~,~\hat{n} = n \sum_{n=0}^{d-1} |n\rangle \langle n|~,
 \label{eu_eqn}
\end{eqnarray}
$\sigma^{x,z}_i$ are Pauli matrices, and $G_i$ is a Gaussian random variable of zero mean and unit variance. The spin-1/2 Hamiltonians $H_{\pm}$ yield static models with both MBL and thermal phases.  The operators $\hat a$ and $\hat n$ play the role of lowering and number operators for the central mode. In this work, we mainly study the case of a central $d$-level system -- a qudit -- for which $\hat a$ lowers the excitation number by one with unit matrix element; this will be compared to photons and central spin-$S$ later in the paper. The qudit levels are split by a bare energy $\Omega$ (with $\hbar=1$) and can be excited through coupling to the spin Hamiltonian $H_-$. The most important parameter for localization is $\Gamma$. The limits $\Gamma=0$ and $1$ represent trivially localized and thermalizing phases, respectively. Other parameters are chosen as $g=0.9045$, $h=0.809$ and $\Omega=3.927$. We expect that our results will be independent of these particular parameters, though we note that MBL is generally favored by large $\Omega$. Furthermore, we will use $d=12$ throughout to approximate $d = \infty$, such that only the lower cutoff on qudit number, $n \geq 0$, plays a role.

\emph{Mobility edge} -- In this model, \cite{NathanMikeCentral} found evidence for an infinite temperature phase transition ($E \approx \mathrm{tr} [H]/\mathrm{tr} [\mathds 1]$) between MBL at small $\Gamma$ and thermalization at large $\Gamma$ upon taking $L\to\infty$ at finite $d/\sqrt{L}$. The transition occurs at $\Gamma_c \approx 0.33$ for $d/\sqrt{L} \gg 1$, which corresponds to the Floquet limit. In this paper, we will study the energy dependence of this transition. In order to obtain initial insight into energy dependence, we utilize the results of the high-frequency expansion \cite{NathanMikeCentral}, rederived in the Supplementary Information for clarity \cite{Supplement}. Physically, the high-frequency expansion (HFE) involves perturbatively eliminating fluctuations of the central mode via a canonical transformation, similar to the Floquet-Magnus expansion \cite{BukovHFE, GoldmanHFE} or Schrieffer-Wolff transformation \cite{BukovMikeAnatolyPRL}. For $d=\infty$, this gives an effective Hamiltonian,
\begin{equation*}
H_{\mathrm{eff}}=  \frac{H_+}{2} - \frac{(H_-)^2}{16 \Omega} |0\rangle \langle 0| + \hat{n} \Omega + O(\Omega^{-2}) 
\end{equation*}
The first term in $H_\mathrm{eff}$ consists of the undriven Hamiltonian $H_+$. The second comes from commutators of the qudit operators, which lead to infinite-range interactions. The leading term is $\sim (H_-)^2/\Omega$ but, importantly, it is only active when the qudit is at its extreme value of $|0\rangle$. For our model, this gives infinite range interactions near the zero energy state, which compete with local interactions in $H_+$ to thermalize the system. Higher order terms will give long-range interactions mediated by states $|1\rangle$, $|2\rangle$, etc., but suppressed by powers of $\Omega^{-1}$.

The HFE suggests the existence of an inverted mobility edge. For large energy, $E/\Omega \gg 1$, for which the qudit number is $n\gg 1$, no infinite-range interactions are produced, and the MBL-delocalized transition is given by that of the locally dressed $H_+$ Hamiltonian with $\Gamma_c \approx 0.33$. For $E\approx 0$ ($n\approx 0$), infinite range interactions compete with $H_+$, generically leading to thermalization.

\begin{figure*} 
\centering
\includegraphics[width=0.8\textwidth]{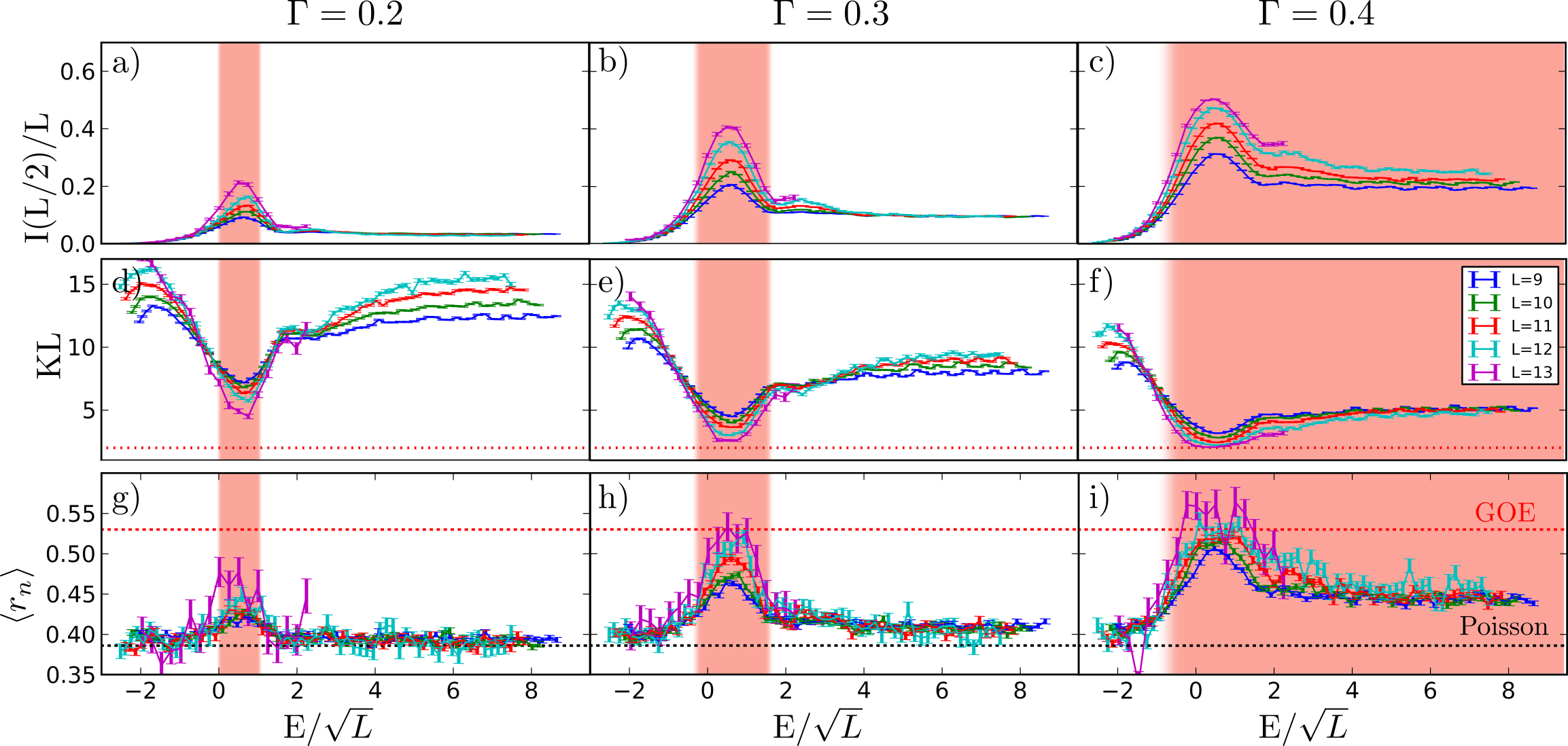}
\caption{ (a-c) Half-system mutual information (I$(L/2)$), (d-f) Kullback-Leibler divergence, and (g-i) level statistic $\braket{r_n}$ as function of energy. The shaded red color indicates the approximate region of delocalization.}
\label{fig:AllData}
\end{figure*}

\emph{Numerics} -- To distinguish the MBL and thermal phases numerically, we first study energy eigenstates of the Hamiltonian \eqref{Hez} using shift-and-invert methods \cite{PietracaprinaShiftInvert} to target $10$ eigenstates near a given energy, up to a maximum system size of $L=13$. 
Thermal systems are expected to follow the rules of random matrix theory, while MBL phases do not. Looking at their energy levels, this implies that thermal eigenstates undergo level repulsion, following Wigner-Dyson level statistics, while MBL eigenstates follow Poisson level statistics with no level repulsion. This is captured by the level spacing statistic \cite{OganesyanHuse}:
\begin{equation}
r_n  = \frac{\mathrm{min}(\delta E_n, \delta E_{n+1})}{\mathrm{max}(\delta E_n, \delta E_{n+1})},  
\end{equation}
where $\delta E_n = E_n - E_{n-1}$ is the gap between ordered eigenenergies $E_n$. For Poisson statistics,  $\langle r_n\rangle = 0.3863 \equiv r_{Pois}$, while for the Gaussian orthogonal ensemble, $\langle r_n \rangle = 0.5307 \equiv r_{GOE}$. Figure \ref{fig:AllData}(g-i) shows the numerically calculated level statistics. At the largest $L$, states near $E=0$, corresponding to the bare energy of the qudit ground state $|0\rangle$, converge toward $r_{GOE}$. At both lower and higher energies, the level statistics appear Poissonian, suggesting that the system is localized. An approximate window for thermalization is sketched in the plots based on where the level statistics start to drift towards the GOE value. Interestingly, a reemergent MBL phase appears at low energies $E < 0$. This is consistent with the high-frequency expansion, which at low enough energies will be dominated by the term $-(H_-)^2 / (16 \Omega)$. While this term has been argued to give infinite-range interactions that compete with short-range interactions, in isolation it shares eigenstates with the local Hamiltonian $H_-$. Therefore, MBL for $E < 0$ apparently comes from the static MBL phase of $H_-$. 

Full convergence to $r_{GOE}$ is difficult to see, particularly for small values of $\Gamma$. Therefore, we turn to the half-system mutual information and Kullback-Leibler divergence. The Kullback-Leibler divergence (KL) measures similarity between eigenstates \cite{LuitzMobilityedge}. For each eigenstate $\ket{n}$, a probability distribution is defined by $p_n(i) = \left| \braket{i|n} \right|^2$, where $\ket{i}$ is an element of the $\sigma^z \otimes \hat n$ basis. For two neighboring energy eigenstates $\ket{n}$ and $\ket{n+1}$, the KL is defined by $\mathrm{KL} = \sum_{i}^{dim(H)} p_n(i) \ln{\frac{p_n(i)}{p_{n+1}(i)}}$. In the MBL phase, this quantity increases linearly with system size $\propto \ln{(\mathrm{dim}(H))}$ because nearby eigenstates are completely uncorrelated. For the thermal phase, one expects $\mathrm{KL}=2$ in the thermodynamic limit from random matrix theory \cite{LuitzMobilityedge}. The energy-dependent KL is shown in Figure \ref{fig:AllData}(d-f). The KL of the thermal phase is notably lower than MBL phase and, for small $\Gamma$, shows an inversion of the finite size dependence; KL increases with system size in the MBL phase and decreases with system size in the delocalized phase. A finite size crossing of the KL gives an approximate location of the delocalized phase, which is seen to increase for increasing $\Gamma$.

Similar behavior is seen in the half-system mutual information (MI) of the energy eigenstates, defined as
\begin{equation} \label{MI}
\mathrm{I(L/2)} \equiv \mathrm{I}(A,B) = \mathcal{S}(A) + \mathcal{S}(B) - \mathcal{S}(A \cup B)   
\end{equation}
where $\mathcal{S}(A)=-\mathrm{tr}[\rho_Aln(\rho_A)]$ is the von Neumann entanglement entropy of subsystem $A$. The system is split into three pieces, as shown in the inset to Figure \ref{fig:PhaseDia}, where $A$ and $B$ correspond to dividing the spin system into halves and $S(A\cup B)=S_q$ is the entanglement entropy of the qudit. Mutual information is chosen to best capture entanglement between the subsystems A and B, which should be area law in the MBL phase and volume law in the delocalized phase \cite{Pal2010}. 
As seen in Figure \ref{fig:AllData}(a-c), mutual information is indeed higher in the delocalized phase, though the apparent super-volume-law scaling is a finite size effect which is expected to go away at larger system sizes \cite{NathanMikeCentral, MCTDH}. Note that the mutual information remains well below its maximal (Page) value of I$(L/2) \to L \ln{2} \approx 0.693 L$, further demonstrating the large finite size effects.

To approach larger system sizes up to $L=18$, we use Krylov time evolution \cite{PietracaprinaShiftInvert}, which is limited to shorter times. For the localized phase, we expect the system to retain memory of its initial state to exponentially long time, resulting in a quick plateau of the mutual information, followed by slow -- potentially logarithmic -- growth \cite{SerbynSlowEntang, BardarsonSlowEntan}. By contrast, ergodic phases should quick reach thermal equilibrium with much larger entanglement. The crossover behavior is more complicated, but physics deep in these phases should be well-approximated by this simple picture. 

\begin{figure}
    \includegraphics[width=0.95\linewidth]{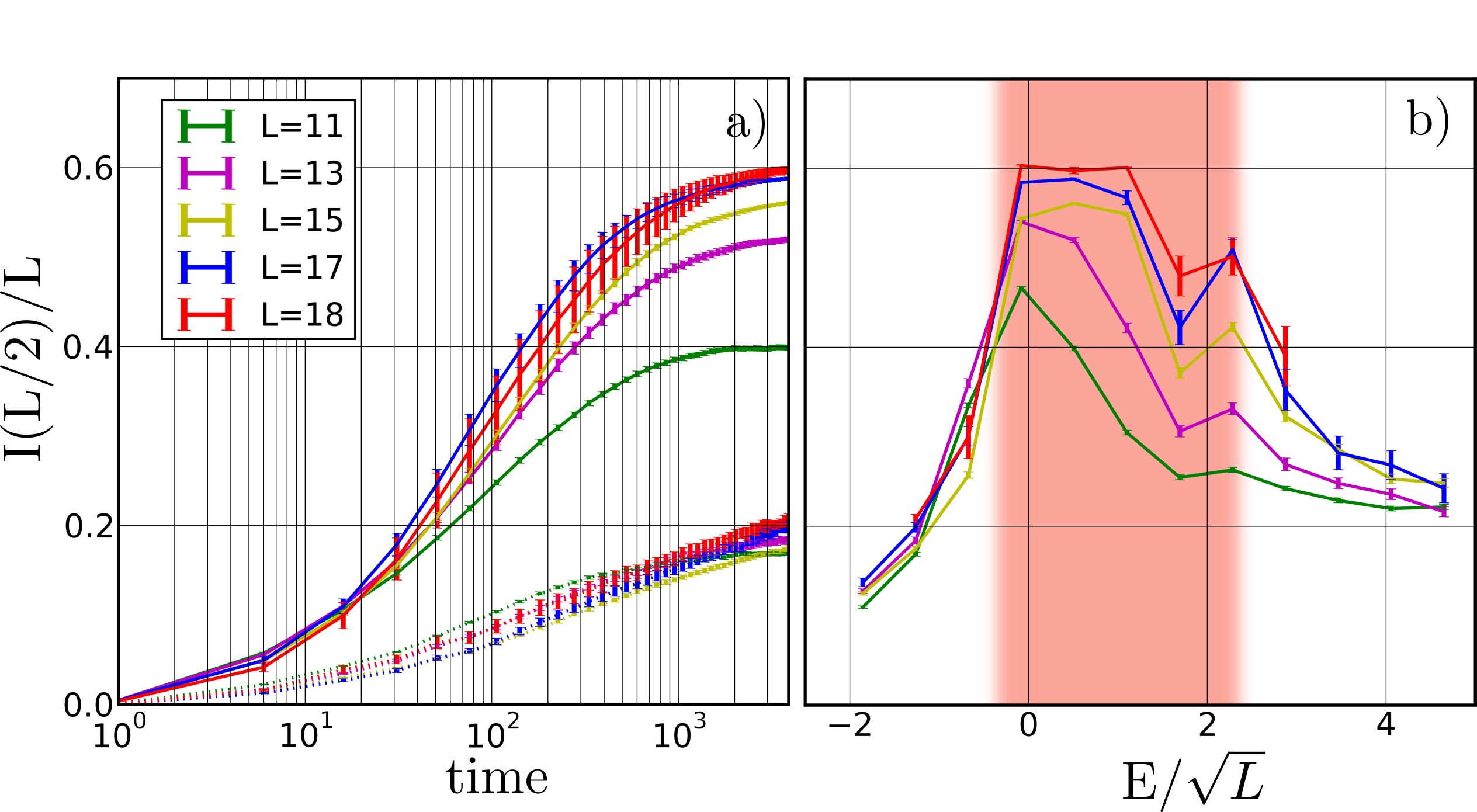}
    \caption{(a) Dynamics of I($L/2$) obtained using  Krylov time evolution starting from a product state. Dashed curves correspond to the MBL regime ($E/\sqrt{L} =-1.26$) and solid lines correspond to the delocalized regime ($E/\sqrt{L} =0.51$). (b) Energy dependence of $I(L/2)$ at late time, $t=3900$. All data are for $\Gamma=0.2$. 
    }
    \label{fig:DynamicEnatn}
\end{figure}

We studied time evolution by preparing initial product states in the $\sigma^z\otimes \hat n$ basis and evolving the wave function using the Krylov method \cite{BrenesKrylov}. Beginning these states within a given energy window of width $\Delta E =0.2 $, Figure \ref{fig:DynamicEnatn} shows energy-resolved mutual information. 
We are not able to obtain data for sufficiently long times to clearly identify a late-time plateau, but points within the MBL and delocalized regions show different trends. For delocalized values of $E/\sqrt{L}=0.5$, the mutual information approaches a plateau value near the theoretical maximum, I($L/2$)$= L\times ln(2)$.  On the other hand, for the states near the ground state and in the middle of the spectrum, data consistent with a logarithmic growth of mutual information is detected, which is suggestive of localization in thermodynamic limit \cite{SerbynSlowEntang, BardarsonSlowEntan}. Taking the instantaneous mutual information at late time, $t=3900$, we observe same trend as the data obtained using energy eigenstates (top panel of Figure  \ref{fig:DynamicEnatn}).

\emph{Discussion} -- 
Our data are consistent with the picture from the high-frequency expansion suggesting an inverted mobility edge for $\Gamma < \Gamma_c$. A concern for this analysis is the fact that the HFE is asymptotic rather than convergent for $L=\infty$ and finite $\Omega$ \cite{WEINBERGPhysReport}. Therefore, we numerically compare the results from the exact numerics to those from the HFE. 
The HFE matches very well in the low energy re-emergent MBL phase and appears to approach the correct answer in the high energy phase, where convergence is expected when the system is localized by standard arguments for Floquet MBL \cite{Lazarides2015,Ponte2015}. Unsurprisingly, the HFE does not converge in the thermal regime, which is a signature of resonant delocalization \cite{Supplement}.

The HFE also shows why inverted mobility edges are more apparent with central qudits or spins than with bosonic modes. In the photonic HFE, the leading long-range interactions become $-(H_-)^2/\Omega$ independent of photon number \cite{Supplement}. Higher order corrections will pick up photon number dependence, but are more difficult to see due the $\Omega^{-r}$ suppression at $r$th order. For a central spin $S$, the relevant commutator is $[S^-,S^+]=-2S^z$, which becomes large at the edges of the spin spectrum (large $|S^z|$), similar to the qudit, and thus will also show an inverted mobility edge as seen in the supplement \cite{Supplement}. In general, the energy-dependence of localization will depend on the manner in which the photon couples to the many-body system, and similar HFEs should enable analysis of the energy dependence.

We can also get some insight from the HFE about the energy at which the MBL-delocalized transition occurs. Within the HFE, the density of states for each qudit level $\ket{n}$ is approximately Gaussian with mean $n\Omega$ and width $\sim J \sqrt L$. 
From the HFE, only states from the $n=0$ branch contribute to the infinite-range thermalizing interactions at leading order. Since the energy window corresponding to $n=0$ extends up to $E_0 \approx J \sqrt{L}$, we postulate that the critical energy will scale similarly: $E_c \propto L^{1/2}$. We are unable to definitively confirm this scaling given our small, finite-size numerics. However, plotting data as a function of $E/\sqrt{L}$ -- as is done throughout the paper -- appears to give better data collapse than plotting as a function of $E$ (see Supplemental material \cite{Supplement}).

Experimentally, a few systems exist in which a non-bosonic central mode is globally coupled to an interacting spin or electron system as required for this physics. A notable example is the recent realization of a cavity QED-like architecture with superconducting qubits playing the role of mirrors \cite{MirhosseiniCavityLike,AlbrechtSubradiant}. The cavity mode is replaced by the dark state manifold of a qubit chain, whose raising and lowering operators satisfy the commutation relations of large spin-$S$) \cite{Supplement}. The size of this spin is controllable by the number of qubits in the chain, hence can be scaled to large values as we use here. Currently, experiments have shown coupling of the dark mode to a single atom-like qubit to simulate cavity QED, but we expect that coupling to a disordered interacting spin chain is practical through conventional superconducting qubit architectures \cite{ChenPedramgmon}. Similar large-spin algebra results for coupling between polaritons in a semiconductor microcavity and spin impurities in the semiconductor, since in certain regimes the polaritons ``inheret'' the non-bosonic commutation relations of their matter component. \cite{QuinteiroPolaritonExp,hartmann2006strongly}. 

Finally, we note that, for generic cavity-atom coupling in conventional cavity QED we also expect an inversion of the mobility edge in certain regimes, as will be detailed in an upcoming paper \cite{RongchunMichael}.

In summary, we have shown that in centrally coupled spin chains, such as those with a central qudit or spin-$S$ in a magnetic field, an inversion of the mobility edge is possible. We postulate that this will be a generic feature of many such models, 
since long-range thermalizing interactions are most strongly induced at the edge of the spectrum where the compactness of the central mode becomes apparent. This phenomology opens up further intriguing questions about localization in such systems with competition between local and global interactions, such as the existence and character of localized bits ($\ell$-bits \cite{Serbynl-bit, HuseOganesyanNandkishorel-bit}). Furthermore, as the energy-dependent phase transition comes from global interactions, it should be in a different class than recent avalanche pictures of the MBL transition \cite{Zhang2016,Goremykina2019,Morningstar2019,Mace2019}.

\emph{Acknowledgments} -- We would like to acknowledge useful discussions with R. Nandkishore, N. Ng, and A. Polkovnikov. This work was performed with support from the National Science Foundation through award number DMR-1945529 and the Welch Foundation through award number AT-2036-20200401. We used the computational resources of the Lonestar 5 cluster operated by the Texas Advanced Computing Center at the University of Texas at Austin and the Ganymede and Topo clusters operated by the University of Texas at Dallas' Cyberinfrastructure \& Research Services Department

\bibliographystyle{ieeetr}

\bibliography{references}

\clearpage
\newpage
\onecolumngrid

\section*{Supplementary Information}

\subsection{More details regarding the high-frequency expansion}

\subsubsection{Non-Floquet derivation of high-frequency expansion}

Consider the following generic cavity/qubit many-body Hamiltonian:
\begin{equation}
H=\Omega\hat{n}+H_{0}+H_{1}(\hat{a}+\hat{a}^{\dagger}),
\label{eq:static_cavity_qubit_hamiltonian}
\end{equation}
where we can either have $\hat{a}$ represent qudit lowering operator,
$\hat{a}_{q}|n\rangle=(1-\delta_{n0})|n-1\rangle$, bosonic annihiliation
operator, $\hat{a}_{b}|n\rangle=\sqrt{n}|n-1\rangle$, or spin lowering
operator, $\hat{a}_{s}|n\rangle=\sqrt{1-n(n-1)/[s(s+1)]}|n-1\rangle$,
where for spins $n=-s,-s+1,\ldots,s$ and otherwise $n=0,1,\ldots$.
We've picked the simplest case where there is only a single harmonic
and $H_{1}$ is real so that it only couples to the real quadrature
of the cavity, but this expansion can be easily amended to treat other
terms. The idea is to do a high-frequency expansion, i.e., a perturbative
expansion around $\Omega=\infty$. Therefore we only want to diagonalize
the $\hat{n}$ operator and don't care about diagonalizing the spin/electron
operators. We will do so by canonical transformation, similar to Schrieffer-Wolff:
\begin{equation*}
H_{eff}=e^{iS}He^{-iS},\;S=\frac{S_{1}}{\Omega}+\frac{S_{2}}{2!\Omega^{2}}+\frac{S_{3}}{3!\Omega^{3}}+\cdots,
\end{equation*}
where each $S_{j}$ is Hermitian. At second order, we can expand the
exponentials and collect terms:
\begin{align*}
H_{\mathrm{eff}} & \approx\left(1+\frac{iS_{1}}{\Omega}+\frac{iS_{2}}{2\Omega^{2}}-\frac{S_{1}^{2}}{2\Omega^{2}}\right)\left(\Omega\hat{n}+H_{0}+H_{1}(\hat{a}+\hat{a}^{\dagger})\right)\left(1-\frac{iS_{1}}{\Omega}-\frac{iS_{2}}{2\Omega^{2}}-\frac{S_{1}^{2}}{2\Omega^{2}}\right)\\
 & \approx\hat{n}\Omega+\left(iS_{1}\hat{n}-i\hat{n}S_{1}+H_{0}+H_{1}(\hat{a}+\hat{a}^{\dagger})\right)+\\
 & \;\;\frac{1}{\Omega}\left(\hat{n}\left(-\frac{iS_{2}}{2}-\frac{S_{1}^{2}}{2}\right)+\left(H_{0}+H_{1}(\hat{a}+\hat{a}^{\dagger})\right)\left(-iS_{1}\right)+\left(iS_{1}\right)\left(\hat{n}\right)\left(-iS_{1}\right)+\left(iS_{1}\right)\left(H_{0}+H_{1}(\hat{a}+\hat{a}^{\dagger})\right)+\left(\frac{iS_{2}}{2}-\frac{S_{1}^{2}}{2}\right)\hat{n}\right)\\
 & \approx\hat{n}\Omega+\left(i\left[S_{1},\hat{n}\right]+H_{0}+H_{1}(\hat{a}+\hat{a}^{\dagger})\right)+\\
 & \;\;\frac{1}{\Omega}\left(i\left[S_{2},\hat{n}\right]-\frac{1}{2}\left(\hat{n}S_{1}^{2}-2S_{1}\hat{n}S_{1}+S_{1}^{2}\hat{n}\right)+i\left[S_{1},H_{0}+H_{1}(\hat{a}+\hat{a}^{\dagger})\right]\right).
\end{align*}

The goal is to select $S_{j}$ order-by-order to cancel off-diagonal
terms of $\hat{n}$. The first order term,
\begin{equation*}
H_{\mathrm{eff}}^{(1)}=i\left[S_{1},\hat{n}\right]+H_{0}+H_{1}(\hat{a}+\hat{a}^{\dagger})
\end{equation*}
 has two off-diagonal terms, $H_{1}\hat{a}$ and $H_{1}\hat{a}^{\dagger}$.
We want to pick $S_{1}$ to cancel these, so the natural ansatz is
$S_{1}=A_{1}\hat{a}+A_{2}\hat{a}^{\dagger}$, where $A_{1,2}$ are
matrices acting on the spins/electrons. For any of the above choices
for $\hat{a}$, we see that 
\begin{equation*}
\left[\hat{a},\hat{n}\right]|n\rangle=nc_{n}|n-1\rangle-c_{n}(n-1)|n-1\rangle=c_{n}|n-1\rangle=\hat{a}|n\rangle,
\end{equation*}
where we define $\hat{a}|n\rangle=c_{n}|n-1\rangle$. Thus $\left[\hat{a},\hat{n}\right]=\hat{a}$.
Similarly, $\left[\hat{a}^{\dagger},\hat{n}\right]=-\hat{a}^{\dagger}$.
Plugging in the ansatz for $S_{1}$, we have 
\begin{equation*}
i\left[S_{1},\hat{n}\right]=iA_{1}\hat{a}-iA_{2}\hat{a}^{\dagger}.
\end{equation*}
Thus, to cancel out the off-diagonal terms, we must choose $A_{1}=iH_{1}$
and $A_{2}=-iH_{1}$, i.e., $S_{1}=iH_{1}(\hat{a}-\hat{a}^{\dagger})$.

One can follow a similar strategy for $S_{2}$, $S_{3}$, etc. with
the natural ansatz $S_{j}=M_{1}\hat{a}+M_{1}^{\dagger}\hat{a}^{\dagger}+M_{2}\hat{a}^{2}+M_{2}^{\dagger}(\hat{a}^{\dagger})^{2}+\ldots+M_{j}\hat{a}^{j}+M_{j}^{\dagger}(\hat{a}^{\dagger})^{j}$.
To obtain the effective Hamiltonian at 2nd order, we simply plug in
the results from 1st order and drop all off-diagonal terms (which
will be accomplished by $S_{2}$).
\begin{align*}
H_{\mathrm{eff}} & \approx\hat{n}\Omega+H_{0}+\frac{1}{\Omega}\left(\frac{H_{1}^{2}}{2}\left(\hat{n}\left(\hat{a}-\hat{a}^{\dagger}\right)^{2}-2\left(\hat{a}-\hat{a}^{\dagger}\right)\hat{n}\left(\hat{a}-\hat{a}^{\dagger}\right)+\left(\hat{a}-\hat{a}^{\dagger}\right)^{2}\hat{n}\right)-\left[H_{1}\left(\hat{a}-\hat{a}^{\dagger}\right),H_{1}(\hat{a}+\hat{a}^{\dagger})\right]\right)_{\mathrm{diag}}\\
 & =\hat{n}\Omega+H_{0}+\frac{H_{1}^{2}}{\Omega}\left(\frac{1}{2}\left(-\hat{n}\left(\hat{a}\hat{a}^{\dagger}+\hat{a}^{\dagger}\hat{a}\right)+2\hat{a}\hat{n}\hat{a}^{\dagger}+2\hat{a}^{\dagger}\hat{n}\hat{a}-\left(\hat{a}\hat{a}^{\dagger}+\hat{a}^{\dagger}\hat{a}\right)\hat{n}\right)-2\left[\hat{a},\hat{a}^{\dagger}\right]\right)\\
 & =\hat{n}\Omega+H_{0}+\frac{H_{1}^{2}}{\Omega}\left(\frac{1}{2}\left(-\left[\cancel{\hat{a}\hat{n}}-\hat{a}\right]\hat{a}^{\dagger}-\left[\cancel{\hat{a}^{\dagger}\hat{n}}+\hat{a}^{\dagger}\right]\hat{a}+\cancel{2\hat{a}\hat{n}\hat{a}^{\dagger}}+\cancel{2\hat{a}^{\dagger}\hat{n}\hat{a}}-\hat{a}\left[\cancel{\hat{n}\hat{a}^{\dagger}}-\hat{a}^{\dagger}\right]-\hat{a}^{\dagger}\left[\cancel{\hat{n}\hat{a}}+\hat{a}\right]\right)-2\left[\hat{a},\hat{a}^{\dagger}\right]\right)\\
 & =\hat{n}\Omega+H_{0}-\frac{H_{1}^{2}}{\Omega}\left[\hat{a},\hat{a}^{\dagger}\right].
\end{align*}
This is the same expression derived in \cite{NathanMikeCentral}
with the notable exception that the harmonic level spacing $\hat{n}\Omega$
remains explicitly present. In that paper it was instead argued to
derive by adiabatic continuation from the rotating frame to the lab
frame. The commutator $\left[\hat{a},\hat{a}^{\dagger}\right]$ is
where the difference between qudit, boson, and spin-$S$ arises. It
is, respectively,
\begin{equation} \label{AllAcommutations}
\begin{aligned}
\left[\hat{a}_{b},\hat{a}_{b}^{\dagger}\right] & =1\\
\left[\hat{a}_{q},\hat{a}_{q}^{\dagger}\right] & =\left(\sum_{n=1}^{d-1}|n-1\rangle\langle n|\right)\left(\sum_{n=1}^{d-1}|n\rangle\langle n-1|\right)-\left(\sum_{n=1}^{d-1}|n\rangle\langle n-1|\right)\left(\sum_{n=1}^{d-1}|n-1\rangle\langle n|\right)=|0\rangle\langle0|-|d-1\rangle\langle d-1|\\
\left[\hat{a}_{s},\hat{a}_{s}^{\dagger}\right] & =\frac{2\hat{n}}{s(s+1)}
\end{aligned}
\end{equation}
While higher order terms can be recovered by a similar procedure,
the Floquet high-frequency expansion in general makes this simpler
by replacing direct solution of canonical perturbation theory by a
solved Floquet problem \cite{GoldmanHFE, BukovKolodrubetz}.

\subsubsection{Floquet derivation of high-frequency expansion}

To obtain higher order terms, we follow the same approach as in \cite{NathanMikeCentral, GoldmanHFE}. Specifically, we go to a rotating frame where the harmonic degrees of freedom in the static Hamiltonian in Eq. \ref{eq:static_cavity_qubit_hamiltonian} are replaced by Floquet time-dependence on the $a$ and $a^\dagger$ terms. Then we have
\begin{equation}
H_{rot}=H_{0}+H_{1}(\hat{a}e^{-i\Omega t}+\hat{a}^{\dagger}e^{-i\Omega t}),
\end{equation}
Then we calculate higher-order terms of the effective Hamiltonian $H_{\mathrm{eff}}$ using van Vleck expansion \cite{GoldmanHFE}:

\begin{equation}\label{HFEFull}
\begin{aligned}
H^\mathrm{rot}_\mathrm{eff} =& \sum_{n=-1}^\infty \frac{1}{\Omega^n} H_\mathrm{eff}^{(n)} \\
H_\mathrm{eff}^{(-1)} =& \hat{n} \Omega\\
H_\mathrm{eff}^{(0)} =& H^{(0)} = H_0\\
H_\mathrm{eff}^{(1)} =& [H^{(1)},H^{(-1)}] = H_1^2[a^\dagger,a]\\
H_\mathrm{eff}^{(2)} =&[[H^{(1)},H^{(0)}], H^{(-1)}]+h.c. = [[H_1,H_0] a^ +,H_1 a]+h.c.\\
=&[[H_1,H_0],H_1] \Big( a^\dagger a + a a^\dagger \Big)\\
H_\mathrm{eff}^{(3)}=&\frac{1}{2}[[[H^{(1)},H^{(0)}],H^{(0)}],H^{(-1)}]+\frac{1}{4}[[H^{(1)},[H^{(1)},H^{(-1)}]],H^{(-1)}]+h.c.\\
=& \frac{1}{2}[[[H_1,H_0],H_0],H_1]\Big( a^\dagger a + a a^\dagger \Big)+\frac{1}{2}H_1^4[[a^\dagger,[a^\dagger,a]],a]\\
\end{aligned}
\end{equation}
In this picture the global mode is decoupled from the spin chain, meaning $[H_\mathrm{eff},\hat n] = 0$ and thus we have an $n$-dependent spin chain Hamiltonian $\langle n | H_\mathrm{eff} | n \rangle$. 

Using this effective Hamiltonian, one can derive the time evolution in the rotating frame $U_{rot}(t)$ using a time-dependent kick operator $iK_\mathrm{eff}(t)$,
\begin{equation}
    U_{rot}(t) = e^{-iK^\mathrm{rot}_\mathrm{eff}(t)} e^{-iH^\mathrm{rot}_\mathrm{eff}t} e^{iK^\mathrm{rot}_\mathrm{eff}(0)}
\end{equation}
where the first few terms of the kick operator are,
\begin{equation}\label{Keff}
\begin{aligned}
iK^\mathrm{rot}_\mathrm{eff}(t) =& \sum_{n=-1}^\infty \frac{1}{\Omega^n}  iK_\mathrm{eff}^{(n)}(t)\\
iK_\mathrm{eff}^{(0)}(t) =& 0\\
iK_\mathrm{eff}^{(1)}(t) =& (e^{i\Omega t}H_1 a^\dagger-h.c.)\\
iK_\mathrm{eff}^{(2)}(t) =&e^{i\Omega t}[H_1 a^\dagger,H_0]-h.c.\\
iK_\mathrm{eff}^{(3)}(t) =&e^{i\Omega t}[[H_1,H_0],H_0]a^\dagger + \frac{e^{i\Omega t}}{4}[[H_1,H_0],H_1](a^\dagger)^2+\frac{2e^{i\Omega t}}{3}H_1^3[a^\dagger,[a^\dagger,a]]-h.c.
\end{aligned}
\end{equation}
Finally, using the rotation operator $e^{-i \hat{n} \Omega t}$, we can go back to the lab frame:
\begin{equation}
e^{iK^\mathrm{lab}_\mathrm{eff}(t)} = e^{-i \hat{n} \Omega t} e^{iK^\mathrm{rot}_\mathrm{eff}(t)}, \quad H^\mathrm{rot}_\mathrm{eff} = H^\mathrm{lab}_\mathrm{eff}
\end{equation}

For comparison with numerical simulations of the full Hamiltonian, Eq. \ref{eq:static_cavity_qubit_hamiltonian}, we use the qudit algebra as in the main text:
\begin{equation}
\hat{a}_q = \sum_{n=1}^{n-1} \ket{n-1}\bra{n}, \quad \hat{n} = \sum_{n=1}^{d-1} n\ket{n}\bra{n},\quad
\left[\hat{a}_{q},\hat{a}_{q}^{\dagger}\right] =|0\rangle\langle0|-|d-1\rangle\langle d-1|.
\end{equation}
Substituting these relations into Eq. \ref{HFEFull}, we can derived $H_{\mathrm{eff}}$ for $d\rightarrow \infty$:
\begin{equation}\label{QuditHeff}
\begin{aligned}
H_\mathrm{eff} =& \hat{n} \Omega +\frac{H_+}{2} + \frac{(H_-)^2}{16\Omega}\Big( |0\rangle\langle0| \Big) + \frac{1}{16\Omega^2}[[H_-,H_+],H_-] \Big( 2-|0\rangle\langle0| \Big) + \\ &\frac{1}{32\Omega^3} \Big( \frac{1}{2}[[[H_-,H_+],H_+],H_-]( 2-|0\rangle\langle0| )+  \frac{1}{512}H_-^4(\ket{0}\bra{0} - \ket{1}\bra{1} ) \Big) + O(\Omega^{-4})
\end{aligned}
\end{equation}
In this expansion, the zeroth order term and all terms with commutators are short-range-interacting with standard MBL-thermal phase transitions. But terms proportional to $H_-^{n+1}/\Omega^n$ for odd $n$ are non-local and introduce all-to-all coupling. In expansion terms we derived here, these non-local terms only show up for the states at extreme of qudit levels, here  $\ket{0}$ and $\ket{1}$, but it also will be present in higher-order terms for other qudit levels. These higher order terms are suppressed by the $\Omega^{-n}$ factors, and thus it can be ignored for large $\Omega$. 

The presence of these all-to-all terms is responsible for delocalizing the states near the edge of the spectrum at $\Gamma < \Gamma_c$. In our study, we use a moderate frequency ($\Omega=5\pi/4$), for which the high-frequency expansion is, at best, asymptotic. For even higher frequencies, the dominant terms at our system size would simply be a time average of the original Hamiltonian, which cannot give rise to new physics. These effect of coupling between number sectors would then become apparent at much larger system sizes, beyond our numerical reach. Lower frequencies, meanwhile, will destroy convergence of the high-frequency expansion. 

\subsubsection{Numerical tests of the high-frequency expansion}

\begin{figure}[h]
    \centering
    \includegraphics[width=1\linewidth]{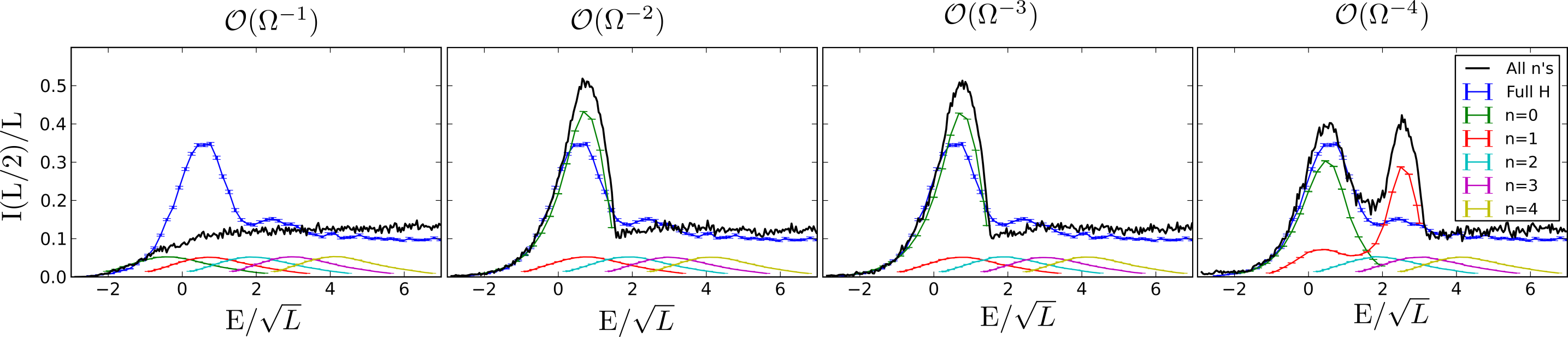}
    \caption{Mutual information I($L/2$) calculated for each global mode level  separately in $H_\mathrm{eff}$. For $E<0$, the major contribution is from $n=0$, which strongly supports our argument that localization of these states is due to the $(H_1)^2$ term. For states near the middle of the spectrum, the sum of all separate levels is given in a dashed black line. The match between actual Hamiltonian and  $H_\mathrm{eff}$ depends on the expansion order. As we include higher-order terms, the localized phase seems to be converging, while the delocalized phase does not. Slow convergence even in the localized phase is expected, but should eventually occur at higher energies were the full Hamiltonian is equivalent to Floquet MBL.}
    \label{Fig_Heff}
\end{figure}

Finally, we numerically verified whether the effective Hamiltonian, Eq. \ref{QuditHeff}, fits the actual data up to order $O(\Omega^{-4})$. In the high-frequency expansion picture, the Hamiltonian is diagonalized within the qudit Hilbert space, giving a block diagonal matrix of size $d\times d$. In this picture, the first block refers to the $n=0$ qudit level, second block to the $n=1$ qudit level, and so on. Each block is individually calculated and then the results are summed over qubit sectors, taking care to apply the kick operator to rotate back to the lab frame. The result is shown in Figure \ref{Fig_Heff}. In all cases, the low energy states of the full Hamiltonian are consistent with the high-frequency expansion, since the $(H_-)^2$ term dominates.
The inclusion of higher-order terms alters blocks at higher $n$, closer to the middle of the spectrum. Since the frequency we have chosen is not particularly large, we find the expected results that a large number of $H_{\mathrm{eff}}$ expansion terms are required to get an acceptable numerical consistency between the full Hamiltonian and expansion terms. We note that convergence is much cleaner in the high-energy MBL phase than the $E \approx 0$ thermal phase -- where the HFE is expected to break down -- but that the MBL has not fully converged by fourth order. We nevertheless expect that the HFE will converge within the MBL phase at sufficiently high order, as this is equivalent to the well-established Floquet MBL phase in the extended zone picture.

\begin{figure}[h]
    \centering
    \includegraphics[width=0.5\linewidth]{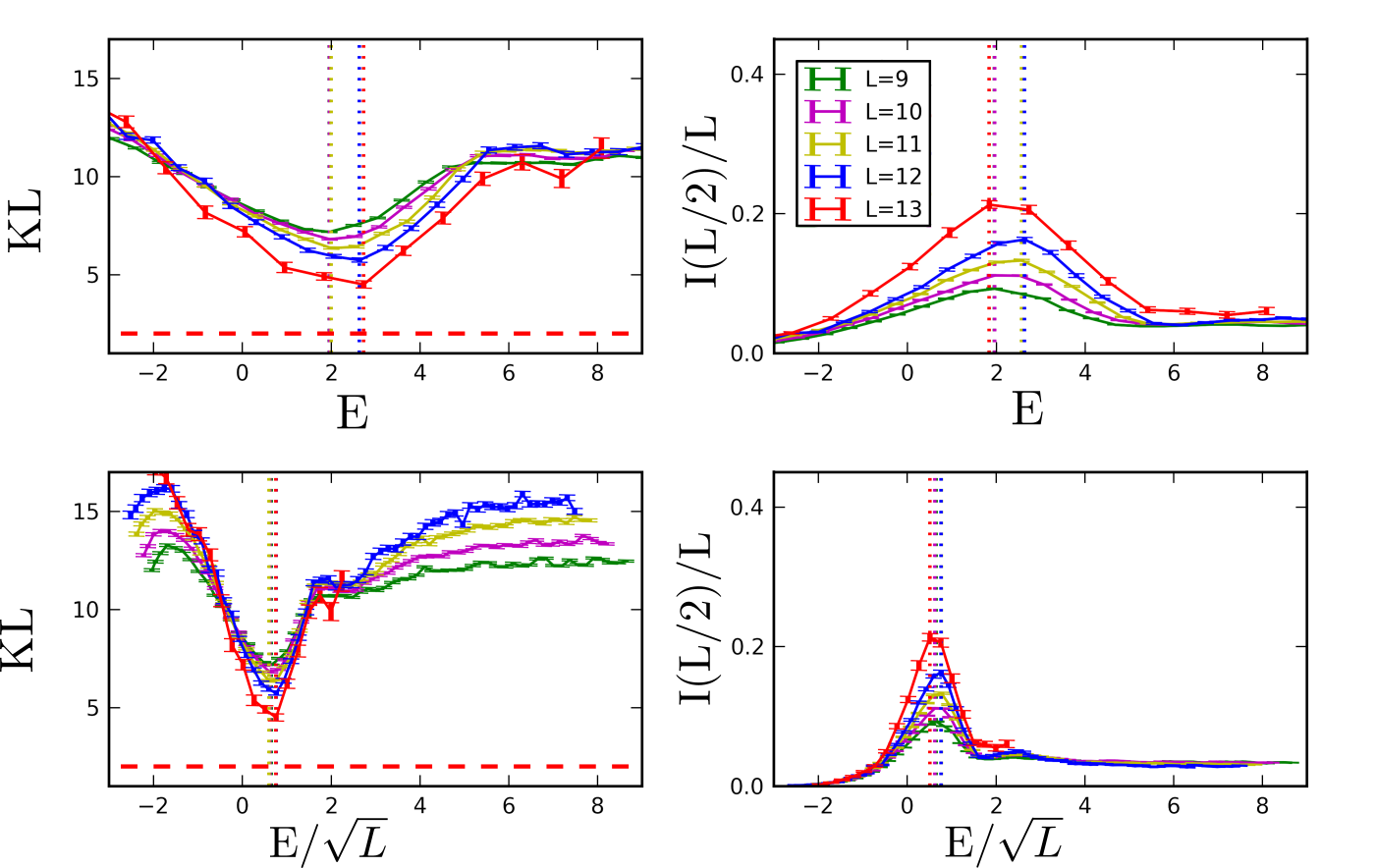}
    \caption{
    Mutual information and KL divergence as a function of either unscaled energy $E$ or scaled energy $E/\sqrt{L}$. All plots are for $\Gamma = 0.2$. Scaling $E/\sqrt{L}$ appears to give better collapse. Dashed lines show the maximum values (minimum values) for mutual information (KL divergence), which seem to converge better for $E/\sqrt{L}$ scaling.} 
    \label{Fig_Escalling}
\end{figure}

Having argued for the consistency of the HFE, we can now use it to predict properties such as the scaling of the transition energy with system size. We use the fact that density of states (DOS) for the short-range interacting spin chain has Gaussian distribution of the form \cite{KeatingDOS},
\begin{equation}
D_n(E) \approx \frac{2^L}{\sqrt{2\pi L}J_{\mathrm{eff}}}\exp \left( -\frac{1}{2}\left( \frac{E-n\Omega}{J_{\mathrm{eff}}\sqrt{L}}\right) ^2\right),
\end{equation}
where we assumed that this distribution is centered at $n\Omega$ for qudit level $n$. This form of DOS should be correct for a finite value of $L$, though as we go to the thermodynamic limit, the non-local terms such as $(H_-)^2$ can modify the density of states at low $n$ \cite{NathanMikeCentral}. Setting aside this potential issue, the standard deviation of the DOS for $n=0$ should set the approximate energy scale of the MBL-delocalized phase transition, since this sets the scale over which these non-local terms compete with local Hamiltonians at higher $n$. This predicts a transition energy $E_c\propto L^\frac{1}{2}$. To test this numerically, we plotted I($L/2$) and KL as a function of either $E$ and $E/\sqrt{L}$. Figure \ref{Fig_Escalling} shows that $E/\sqrt{L}$ scaling gives a better convergence for the maximum (minimum) value of I($L/2$) (KL), though we are unable to state anything conclusively for such small system size.

\subsection{Localization with large-$s$ central spin}

While central qudits provide a numerically tractable test of localization with a central mode, it is not experimentally clear how to realize such a system. Two main types of central mode occur naturally: central bosonic modes, as in cavity QED, or central spins. In this section, we discuss and present data for localization in the presence of a large-$s$ central spin.

For the central spin mode, where $\hat a_s \sim S_-$ and $\hat a \sim S_+$, the algebra reads
\begin{equation}\label{LargeSoperators}
\begin{split}
\hat{a}_s^\dagger &= \frac{1}{\sqrt{s(s+1)}}\sum_{n=-s+1}^{s} \sqrt{s(s+1)-n(n+1)} \ket{n}\bra{n-1},\\ \hat{a}_s &= \frac{1}{\sqrt{s(s+1)}} \sum_{n=-s+1}^{s} \sqrt{s(s+1)-n(n-1)} \ket{n-1}\bra{n}, \\ \hat{n} &=  \sum_{n=-s}^{s}n \ket{n}\bra{n} \propto S_z, \quad \left[\hat{a}_{s},\hat{a}_{s}^{\dagger}\right]  = \frac{2\hat{n}}{s(s+1)}
\end{split}
\end{equation}
Note that the normalization of $\hat a_s$ is chosen such that matrix elements are equal to $1$ near the center of the spectrum (state $|0\rangle$), so that this part of the spectrum reproduces the Floquet extended zone picture. The high-frequency expansion for large central spin is
\begin{eqnarray} 
H^{\mathrm{eff}}_{\mathrm{rot}} &=& \hat{n}\Omega+H_0 + \frac{(H_1)^2}{\Omega}[\hat{a}_{s}^{\dagger},\hat{a}_{s}]+\frac{1}{2\Omega^2}\Big([[H_1,H_0]\hat{a}_{s}^{\dagger},H_1 \hat \hat{a}_{s}]+h.c.\Big)+\dots \nonumber\\
&=& \hat{n}\Omega+ H_0 +  \frac{(H_1)^2}{s(s+1)\Omega}   2\hat{n} + \frac{1}{s(s+1)\Omega^2} [[H_1,H_0],H_1] \Big( \hat{a}_{s}^{\dagger}\hat{a}_{s} + \hat{a}_{s} \hat{a}_{s}^{\dagger} \Big) + \dots
\label{eq:H_eff_central_spin}
\end{eqnarray}

\begin{figure}
    \centering
    \includegraphics[width=0.84\linewidth]{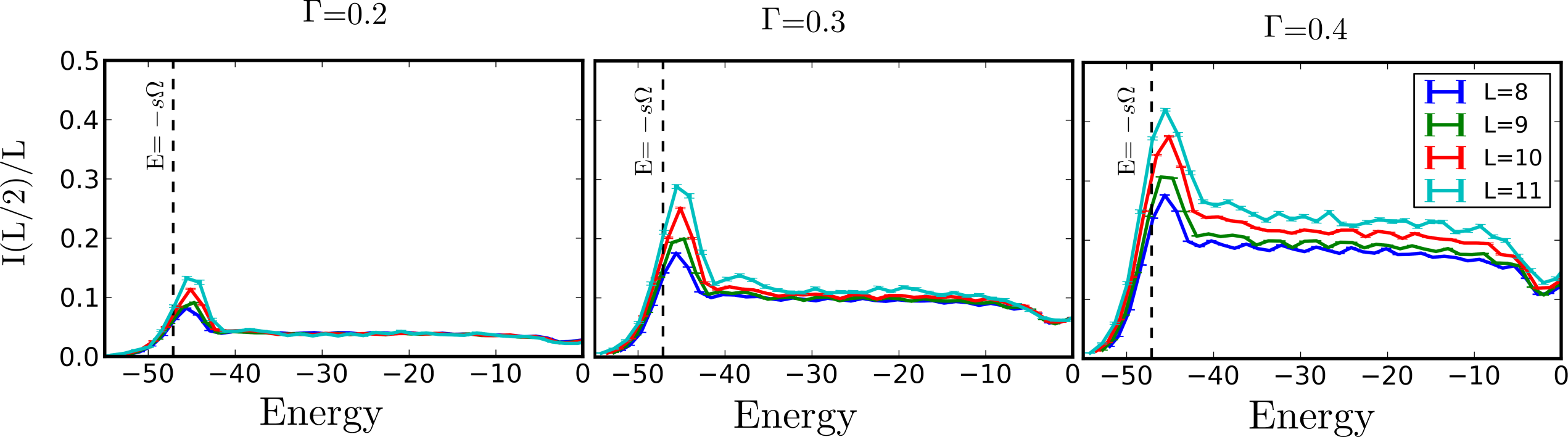}
    \caption{Mutual information as a function of energy using central spin-$s$ with $s=12$. 
    }
    \label{Fig_MISpin}
\end{figure}

To observe the mobility edge numerically for large central spin, we need to find an appropriate regime where the magnitude of the third term in Eq. \ref{eq:H_eff_central_spin} is comparable to that of the second term. Near the low energy states, $\hat n \approx -s$, we can write
\begin{equation}
H_0 +  \frac{(H_1)^2}{s(s+1)\Omega}   2\hat{n} \approx H_0 -  2\frac{(H_1)^2}{(s+1)\Omega}  \approx H_0 -  2H_1 \left[ \frac{(H_1)}{(s+1)\Omega} \right].
\label{eq:expansion_near_edge}
\end{equation}
Since $H_1$ is a local Hamiltonian with extensive energy variance, the relevant magnitude is the standard deviation over energy eigenstates, $\sigma(\langle H_1 \rangle)$, which is proportional to $\sqrt{L}$. Therefore, to make the term in square brackets in Eq. \ref{eq:expansion_near_edge} of order unity, we need $s \apprge \sqrt{L}$. We therefore used moderate system sizes of $L \in [8,9,10,11]$ with $s=12$. The Hamiltonian is identical to that used for the central qudit in the main text with ladder operators $\hat{a}_s$ and $\hat{a}_s^\dagger$ from Eq. \ref{LargeSoperators}. The results are shown in Figure \ref{Fig_MISpin}. Clearly, the mutual information behaves similarly to the mutual information obtained with the central qudit (main text, Figure 2). The mutual information near $E=-s\Omega$ (minimal $\hat n$) is much larger than near the middle or edges of the spectrum. There is a region near zero energy which shows an unexpected decrease in mutual information, suggesting stronger localization. As of yet, we are unable to explain this result, though we note that it occurs near the energy $E=0 \implies n=0$ where the long-range term in the HFE, $(H_-)^2 \hat n / \Omega$, vanishes.

These results suggest that an inverted mobility edge is possible for the experimentally relevant case of large central spin. We note briefly that similar results can be obtained as a side effect of coupling bosonic modes in a different way than in this paper, such as via the vector potential:
\begin{equation*}
    H_\mathrm{int} = \sum_{<jk>} J_{jk} e^{i(\hat a + \hat a^\dagger)C_{jk}} c_j^\dagger c_k.
\end{equation*}
The important difference here is that commutators of the raising and lowering operators enter differently into the high-frequency expansion. This will be described in more detail in an upcoming paper.

In this paper, we focus on central spin-s and qudits. Therefore, in the next section, we show that dark state of cavity-like systems \cite{MirhosseiniCavityLike} have the requisite central spin-$s$ algebra.

\subsection{Collective dark-states with large spin algebra}\label{ProofCollectiveLargespin}

In \cite{MirhosseiniCavityLike}, the authors demonstrate a cavity QED-like architecture in which multiple ``cavity'' qubits are coupled together to form a collective dark state manifold, which in turn couples to a single ``probe'' qubit. While the dark state plays a role similar to the bosonic cavity in cavity QED, here we show that its commutation relations are actually those of a large spin-$s$, with the number of levels $2s+1$ set by the number of cavity qubits.

The collective dark states defined in \cite{MirhosseiniCavityLike} has lowering and raising operators of the form,
\begin{equation}\label{DarkState}
\hat{S}_D = 1/\sqrt{N} \sum_{m>0} (\hat{\sigma}^m_{ge} + \hat{\sigma}^{-m}_{ge}) (-1)^m,
\end{equation}
where $N$, an even number, is the total number of qubits evenly installed about the probe qudit, and $m$ runs up to $N/2$. Here $\hat{\sigma}^m_{ge} = \ket{g_m} \bra{e_m}$ where $g$ and $e$ are referring to the ground and excited states of the $m$th qubit respectively. 
Using Eq. \ref{DarkState}, the commutation relation of raising and lowering operators becomes,
\begin{equation}
\begin{split}
\left[\hat{S}_D,\hat{S}^{\dagger}_D\right] & = 1/N \sum _{m,n>0} \left[\hat{\sigma}^m_{ge} + \hat{\sigma}^{-m}_{ge} ,\hat{\sigma}^{\dagger n}_{ge} + \hat{\sigma}^{\dagger -n}_{ge}\right] (-1)^{m+n}\\
& = 1/N \sum _{m,n>0} \left[\ket{g_m} \bra{e_m} + \ket{g_{-m}} \bra{e_{-m}},  \ket{e_n} \bra{g_n} + \ket{e_{-n}} \bra{g_{-n}}\right](-1)^{m+n}
\end{split}
\end{equation}
Since $\braket{g_m|g_{-n}}=\braket{e_m|e_{-n}}=0$ for any $m$ and $n$, we can write
\begin{equation} \label{Commut2}
\begin{split}
& \left[\hat{S}_D,\hat{S}^{\dagger}_D\right] = 1/N \sum _{m,n>0} \Big( \left[\ket{g_m} \bra{e_m} ,  \ket{e_n} \bra{g_n}\right] + 
 \left[\ket{g_{-m}} \bra{e_{-m}} ,  \ket{e_{-n}} \bra{g_{-n}}\right] \Big) (-1)^{m+n}
\end{split}
\end{equation}
We also have
\begin{equation}
\begin{split}
\left[\ket{g_m} \bra{e_m} ,  \ket{e_n} \bra{g_n}\right] =& \ket{g_m} \bra{e_m} \ket{e_n} \bra{g_n} - \ket{e_n} \bra{g_n} \ket{g_m} \bra{e_m} \\
=& \ket{g_m} \bra{g_n} \delta_{m,n} - \ket{e_n} \bra{e_m} \delta_{m,n}
\end{split}
\end{equation}
Finally, we can write Eq. \ref{Commut2} as
\begin{equation} 
\begin{split}
& \left[\hat{S}_D,\hat{S}^{\dagger}_D\right] = 1/N \sum _{m} \biggl( \ket{g_m} \bra{g_m} - \ket{e_m} \bra{e_m} + 
\ket{g_{-m}} \bra{g_{-n}} - \ket{e_{-n}} \bra{e_{-m}}  \biggr) (-1)^{2m}
\end{split}
\end{equation}
$(-1)^{2m} = 1$. Changing the summation range,
\begin{equation} \label{FinalCommutation}
\begin{split}
&\left[\hat{S}_D,\hat{S}^{\dagger}_D\right] = 1/N \sum _{-\frac{N}{2}<m<\frac{N}{2}, m\neq0} \biggl( \ket{g_m} \bra{g_m} - \ket{e_m} \bra{e_m}  \biggr)
\end{split}
\end{equation}
In this setup, the Hilbert space size is $2\times(N/2)+1$ as for a single spins. For spin we have $[\hat{S}_+, \hat{S}_-]=2\hat{S}_z$, which clearly is equivalent to Eq. \ref{FinalCommutation} up to an overall factor in defining the raising/lowering operators.

\end{document}